# Plasmon Confinement by Carrier Density Modulation in Graphene

Ngoc Han Tu, Makoto Takamura, Yui Ogawa, Satoru Suzuki[†], and Norio Kumada

*NTT Basic Research Laboratories, NTT Corporation, 3-1 Morionosato-Wakamiya, Atsugi 243-0198, Japan*

[†]Present address: University of Hyogo

We investigate plasmon resonances in graphene with periodic carrier density modulation. The period is 8 μm, and each period consists of 1.7- and 6.3-μm-wide ribbons with different density. Using terahertz spectroscopy, we show two plasmon modes with their electric field mostly localized in the 1.7- or 6.3-μm-wide ribbon arrays. We also show that plasmons are excited only in one of the micro-ribbon arrays when the Fermi energy of the other micro-ribbon array is set close to the charge neutrality point. These results indicate that plasmons can be confined by the carrier density modulation.

Graphene plasmons are attracting much attention for fundamental research and photonic device applications in the frequency range from terahertz to mid-infrared due to their striking electrical and optical properties [1-4]. Compared to the surface plasmons in noble metals, graphene plasmons have unprecedented advantages such as extremely strong confinement, relatively low loss, and electrical tunability [5-9]. These features enable to develop graphene-based plasmonic devices such as sensors [10], modulators [11], absorbers [12] and oscillators [13]. Moreover, it has been theoretically proposed that the electrical tunability can provide a versatile platform for dynamically manipulating the plasmon propagation: by precisely tuning the conductivity distribution across a homogeneous graphene sheet, transformation optics devices [14, 15] with their properties variable in real time using electrical gating, which offers the possibility for developing the plasmonic circuitry for information processing, can be formed. On the other hand, plasmonic crystal with a tunable band structure can also be realized by periodically modulating the conductivity [16, 17]. However, effects of the spatial conductivity modulation on graphene plasmons have been poorly explored experimentally.

In this work, we report plasmon resonances in graphene with periodically modulated conductivity in THz range. The conductivity is spatially modulated by chemical doping from a micro-ribbon array of a self-assembled monolayer (SAM) of organosilane formed between graphene and a $SiO_2$/Si substrate. In addition, a global back gate is used to tune the conductivity by shifting the carrier density. When the polarization of the incident light is perpendicular to the micro-ribbon pattern, the extinction spectrum shows two peaks. From the dependence of the peak frequencies on the electrostatic doping from the back gate, we show that the two peaks come from plasmon modes mostly localized in the graphene/SAM and graphene/$SiO_2$ micro-ribbon arrays. We also show that plasmons can be selectively excited in one of the arrays by setting the back gate bias to the charge neutrality point (CNP) of the other array. These results demonstrate that plasmons can be confined by micro-ribbons defined by the conductivity modulation not by etching, and micro-ribbons in which plasmons are excited can be selected by electrical gating. These results will stimulate the development of transformation optics and plasmonic band engineering in graphene.

We used 3-amino-propyltrietthoxysilance (APTES) to form a SAM. The SAM on the $SiO_2$/Si substrate was patterned in an array of 1.7-μm-wide ribbons spaced by 6.3 μm. Homogeneous monolayer graphene grown by chemical vapor deposition was transferred to the patterned substrate [Fig. 1] using established wet transfer processes [18]. Amine functional groups in APTES donate their lone pair of electrons to carbon atoms in graphene [19, 20], inducing the carrier density modulation.

The thickness of the SAM is only a few nanometers, and the plasmon reflection by the height difference at its boundary is negligible [21]. In addition, it is insulating, hardly screen the plasmon electric field. These advantages of the SAM allow us to investigate the effects of carrier density modulation without their being influenced by other mechanisms of plasmon reflection. By applying a bias to the Si back gate, the carrier density can be shifted, while the density difference between the regions with and without the SAM remains fixed.

To probe plasmon excitations, we carried out transmission spectroscopy using a polarized Fourier transform infrared (FTIR) spectrometer with the frequency range between 1.3 and 20 THz. All the measurements were conducted at room temperature. Before every measurement, samples were kept in vacuum (~$10^{-4}$ Pa) overnight and then heated at 80°C for 30 min to exclude effects of moisture.

The two-terminal resistance $R$ of the sample as a function of back gate voltage $V_G$ shows two maxima [Fig. 2(a)], which is consistent with the density modulation. The CNPs for the graphene/SiO$_2$ and graphene/SAM regions are located at $V_{CNP}$ = 55 and 28 V, respectively. Positive $V_{CNP}$ indicates that carriers at $V_G = 0$ V are holes. The carrier densities at $V_G = 0$ V in the graphene/SiO$_2$ and graphene/SAM regions are estimated to be $n = 4.1 \times 10^{12}$ cm$^{-2}$ and $2.1 \times 10^{12}$ cm$^{-2}$, respectively, from the distance to $V_{CNP}$ and the gate capacitance of 115 aF/μm$^2$. Figures 2(b) and (c) show the extinction spectra $1 - T/T_{ref}$ for the incident light polarized parallel $E_\parallel$ and perpendicular $E_\perp$ to the SAM ribbon pattern, respectively, where $T$ and $T_{ref}$ are the transmissions through the sample and the bare substrate, respectively. The behavior for $E_\parallel$ and $E_\perp$ is quite different, indicating that anisotropy of the light-plasmon coupling is introduced into continuous graphene by the carrier density modulation.

For $E_\parallel$, dynamical response of charge carriers is similar to that of free electrons, and the spectrum is well described by the semiclassical Drude model [6]

$$\sigma(\omega) = \frac{iD}{\pi} \frac{1}{\omega + i\Gamma_D}, \quad (1)$$

with a Drude weight $D\hbar^2/e^2 = 0.13$ eV and a scattering rate $\Gamma_D = 2.4$ THz. For comparison, we also obtained $(D\hbar^2/e^2, \Gamma_D) = (0.14, 2.3)$ and $(0.08, 2.0)$ for homogeneous graphene on SiO$_2$ and the SAM, respectively [inset of Fig. 2(b)]. The scattering rates are similar among the three samples, indicating that the SAM formation does not degrade the graphene quality. The difference in the Drude weight is due to the difference in the average carrier density [22].

For $E_\perp$, on the other hand, the spectrum shows two peaks. To explain the extinction spectrum, we recall the plasmon resonance in a graphene micro-ribbon array fabricated by etching [6]. The resonance frequency in micro-ribbons is given by

$$\omega_p = \sqrt{\frac{2e^2 E_F}{\epsilon \hbar^2} q}, \tag{2}$$

where $q = \pi/W$ is the wavenumber, $W$ is the ribbon width, $\epsilon$ is the effective permittivity, and $E_F = \hbar v_F \sqrt{\pi |n|}$ is the Fermi energy. The line shape of the spectrum is $\sigma(\omega) = iD\omega/[\pi(\omega^2 - \omega_p^2 + i\omega\Gamma_p)]$, where $\Gamma_p$ is the plasmon scattering rate. Based on a simple assumption that the observed two peaks come from two independent plasmon resonances, we fitted the spectrum by the linear combination of the two resonance functions:

$$\sigma(\omega) = \alpha \frac{\omega}{\omega^2 - \omega_{p1}^2 + i\omega\Gamma_{p1}} + \beta \frac{\omega}{\omega^2 - \omega_{p2}^2 + i\omega\Gamma_{p2}} \tag{3}$$

where α and β are coefficients. The thick solid line in Fig. 2(c) represents the fitting result, and the thin dashed and solid lines are the contributions of the $\omega_{p1}$ and $\omega_{p2}$ modes, respectively. The fitting satisfactory reproduces the spectrum. The resonance frequencies $\omega_{p1} = 2.8$ THz and $\omega_{p2} = 4.3$ THz obtained by the fitting well agree with the values 2.6 THz and 4.1 THz calculated by Eq. (2) using the parameters for the graphene/SiO$_2$ and graphene/SAM micro-ribbons $(W, n) = (6.3\ \mu m, 4.1 \times 10^{12}\ cm^{-2})$ and $(1.7\ \mu m, 2.1 \times 10^{12}\ cm^{-2})$, respectively, with $\epsilon = 3$. These results indicate that the $\omega_{p1}$ and $\omega_{p2}$ modes have similar electric field distribution to plasmons confined in the 6.3- and 1.7-μm-wide ribbons, respectively. These observations verify that the carrier doping alters the conductivity and thereby induces plasmon reflections at the boundaries between graphene/SiO$_2$ and graphene/SAM regions.

To demonstrate the tunability of plasmon properties, we carried out similar measurements for several values of $V_G$. As shown in Fig. 3, the frequencies of the resonance peaks depend on $V_G$. Particularly, at $V_G = 25$ and 60 V, the spectra show only a single peak within the experimental range. These features can be explained by the change in $E_F$, which can be deduced by the electronic transport data in Fig. 2(a). At $V_G = -20$ V, $E_F$ is tuned to the higher hole-doping regime, and the positions of the two resonance peaks shift to the higher frequency than those of $V_G = 0$ V. This is due to the $E_F$ dependence of the plasmon dispersion [Eq. (2)]. At $V_G = 25$ V, $E_F$ is close to the CNP for the graphene/SAM region. Since plasmons cannot be excited when the system is close to the CNP [6], the observed single resonance peak comes from plasmons confined in the graphene/SiO$_2$ micro-ribbons. The same discussion holds for $V_G = 60$ V, where $E_F$ is close to the CNP for the graphene/SiO$_2$ region and plasmons are confined in the graphene/SAM micro-ribbons. At an intermediate gate bias ($V_G = 45$ V), the graphene/SAM region is electron-doped, while the graphene/SiO$_2$ region is hole-doped. The appearance of the resonance peaks indicates that plasmons are reflected by p-n junctions. At $V_G = 100$ V, the system enters the highly electron-doping regime. In this regime, in contrast to the hole-doping regime, the carrier density in the graphene/SAM region is larger than that in the graphene/SiO$_2$ region. The combination of higher carrier density and narrower ribbon width leads to a higher resonance frequency and vice versa, giving rise to larger splitting of the resonance peaks. Note that the variation of the peak height and width is due to carrier density dependence of Drude weight and scattering rate.

Figure 4 summarizes plasmon resonance frequencies obtained by the fitting using Eq. (3). The frequencies as a function of $V_G$ show that $\omega_{p1}$ and $\omega_{p2}$ decrease as $V_G$ is varied towards $V_{CNP}$ for the graphene/SiO$_2$ and graphene/SAM regions, respectively [Fig. 4(a)]. The frequencies as a function of $|n|^{1/4}$ can be fit by linear functions [Fig. 4(b)], which is characteristic of Dirac fermion systems [23]. The lines in Fig. 4(b) correspond to Eq. (2) for the graphene/SiO$_2$ ($W = 6.3$ μm) and graphene/SAM ($W = 1.7$ μm) micro-ribbons. These results show the electrical tunability of the plasmon frequency in the carrier density modulated system. Furthermore, by selecting the frequency and $V_G$, the spatial distribution of the electromagnetic field associated with plasmons can be tailored.

Finally, we discuss the deviation from the model based on Eq. (3). The model assumes that the system can be treated as independent graphene/SiO$_2$ and graphene/SAM micro-ribbon arrays. Although this simple model provides intuitive interpretation of the observed resonance frequencies, it fails to fit the minimum between the two peaks—as shown in Fig. 3, the fitting always gives a shallower minimum. This implies that the plasmon transmissivity across the boundaries of the micro-ribbons is finite (and the reflectivity is less than unity) and that a minigap in the plasmon dispersion is formed by the periodic carrier density modulation. We suggest that the minimum in the spectra corresponds to the minigap at $q = 2\pi/a$ [16], where $a = 8$ μm is the periodicity. To elaborate this, it is necessary to investigate the spectra while controlling the densities of the two micro-ribbon arrays independently.

In summary, we investigated plasmon excitations under periodic carrier density modulation in graphene. The density modulation was created by forming a SAM micro-ribbon array on a $SiO_2$/Si substrate. Using THz spectroscopy, we observed two plasmonic resonances. From the resonance frequency and its carrier density dependence, we showed that the observed resonances come from plasmon modes mostly localized in the graphene/$SiO_2$ and graphene/SAM micro-ribbon arrays. Particularly, when the Fermi energy of one of the micro-ribbon arrays is tuned to the CNP, plasmons are confined in the other array. This indicates that plasmons can be confined by carrier density modulation, and furthermore, the region in which plasmons are excited can be selected by electrical gating. We also suggested that a minigap is formed in the plasmon dispersion. Our results are important step towards the development of transformation optics and plasmonic band engineering based on spatial control of the conductivity.

**Acknowledgements**
We thank K. Sasaki, M. Hashisaka, and K. Muraki for valuable discussions. This work was supported in part by JSPS KAKENHI (JP16H03835, JP16H06361).

FIG. 1. (a) Optical micrograph of graphene transferred onto a $SiO_2$/ Si substrate covered with a periodic SAM micro-ribbon array. One of the SAM ribbons is highlighted by the dashed lines. (b) Schematic (not drawn to scale) of the experimental setup.

FIG. 2 (a) Two-terminal resistance $R$ as a function of $V_G$. Arrows indicate $V_{CNP}$ for the graphene/$SiO_2$ (55 V) and graphene/SAM (28 V) regions estimated by two-peak fitting (not shown). (b) and (c) Extinction spectra for $V_G = 0$ V (open circles) for $E_\parallel$ and $E_\perp$, respectively. Thick lines are results of fitting. Thin solid and dashed lines in (c) represent contributions of the graphene/SAM and graphene/$SiO_2$ regions, respectively. Inset of (b) shows the spectra for homogeneous graphene on the $SiO_2$ and SAM.

FIG. 3 Extinction spectra for several values of $V_G$ (open circles) for $E_\perp$. Thick lines are results of fitting. Thin solid and dashed lines represent contributions of the graphene/SAM and graphene/$SiO_2$ regions, respectively. Traces are offset for clarity.

FIG. 4 Plasmon resonance frequencies as a function of $V_G$ (a) and $|n|^{1/4}$ (b). Dotted lines in (a) and (b) represent Eq. (2) with $W = 6.3$ and 1.7 μm.

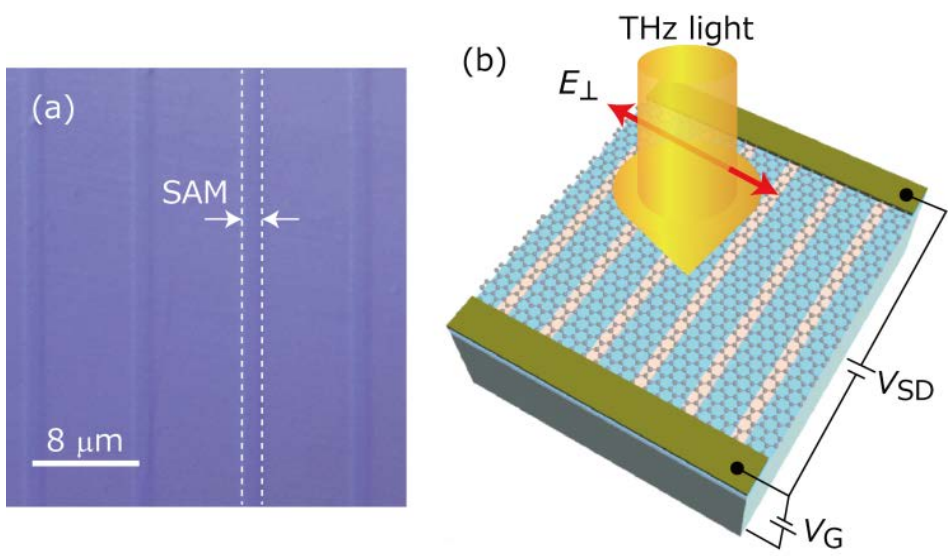

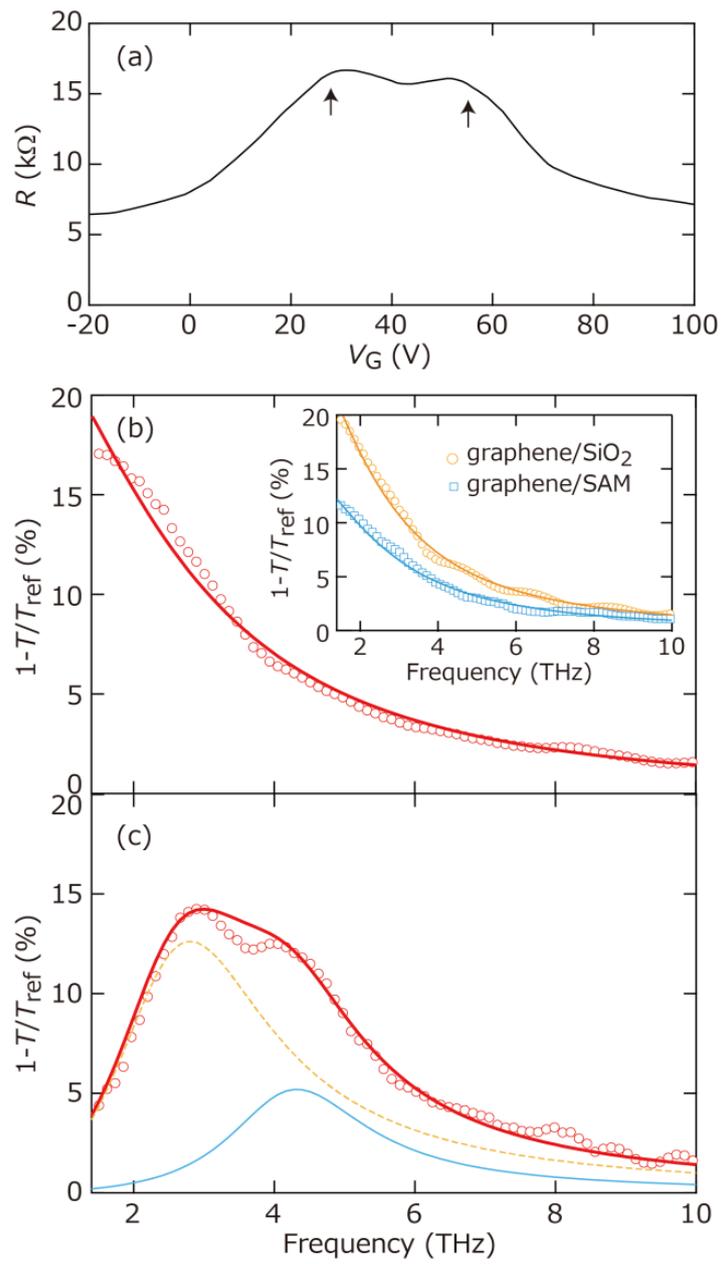

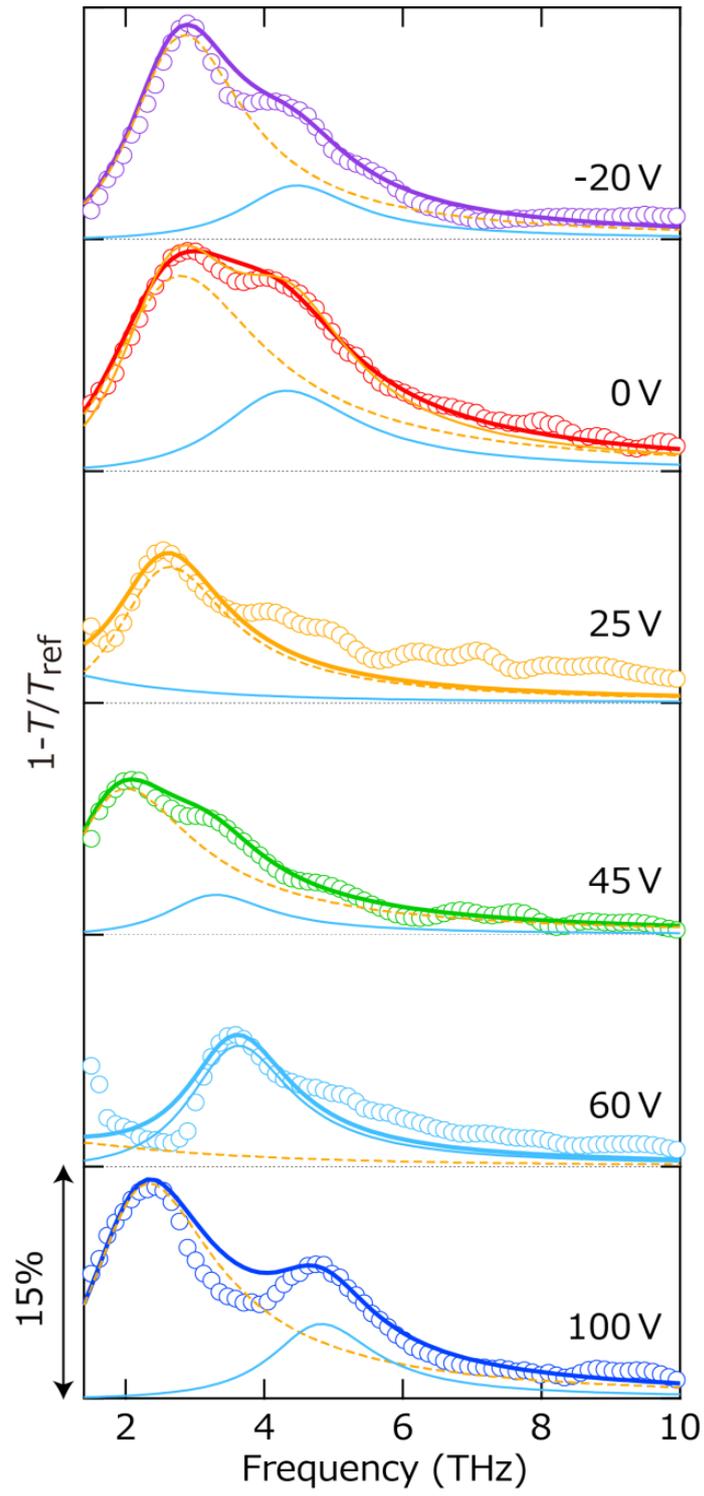

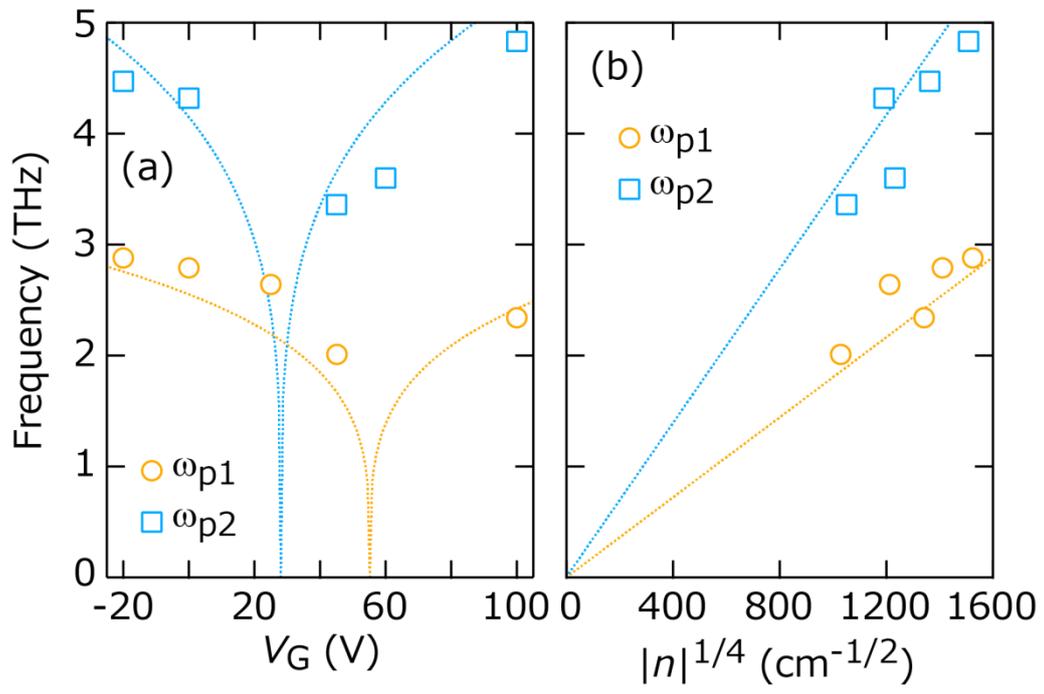

13